\documentclass[pre,aps,showpacs,twocolumn,preprintnumbers,amsmath,amssymb,superscriptaddress]{revtex4}

\usepackage{graphicx}% Include figure files
\usepackage{dcolumn}% Align table columns on decimal point
\usepackage{bm}% bold math

\begin{document}

%\preprint{}

\title{Modeling the interaction of DNA with alternating fields}

\author{A. E. Bergues-Pupo}

\affiliation{Departamento de F\'isica, Universidad de Oriente, 90500 Santiago de Cuba, Cuba}

\affiliation{Dpto. de F\'isica de la Materia Condensada,
Universidad de Zaragoza. 50009 Zaragoza, Spain}

\author{J. M. Bergues}

\affiliation{Escuela Polit\'{e}cnica Superior y Facultad de Ciencias de la Salud, Universidad San Jorge, 50830 Villanueva de G\'{a}llego, Zaragoza, Spain}

\author{F. Falo}

\affiliation{Dpto. de F\'isica de la Materia Condensada,
Universidad de Zaragoza. 50009 Zaragoza, Spain}

\affiliation{Instituto de Biocomputaci\'on y F\'isica de Sistemas Complejos (BIFI), Universidad de Zaragoza, 50009 Zaragoza, Spain}

\date{\today}

\begin{abstract}
We study the influence of a THz field on thermal properties of DNA molecules.  A Peyrard-Bishop-Dauxois model with the inclusion of a solvent interaction term is considered. The THz field is included as a sinusoidal driven force in the equation of motion. We show how under certain field and system parameters, melting transition and bubble formation are modified.

\end{abstract}


\maketitle

\section{Introduction}
Terahertz (THz) technology and science have spread with ever-growing applications in military and security systems, medicine, biology, and researches. To exemplify: security screening at airports \cite{R2,R3}, shipment inspection \cite{R4}, identification of concealed explosives, drugs and weapons \cite{R3,R5,R6}, cancer and burn diagnosis \cite{R7,R8,R9,R10,R11,R12}, and in spectroscopy \cite{R7,R13,R14}. Thus, knowing the effects of THz radiation is critical for different scientific and technological purposes.

Despite the presence of research on the biological effects of THz radiation \cite{R15}, there are still many controversies. In addition, the emergence of strong sources of THz radiation may contribute to the resolution of controversies over the mechanism of biological organization \cite{R16}. The potential of this perspective depends on the development of sophisticated pump-probe and multidimensional experimental techniques and the study of biological systems in the controlled environments necessary for their maintenance and viability \cite{R17}.

In order to assess the possible genotoxicity of THz radiation with biological materials, in the framework of the THz-BRIGDE project \cite{R18}, various studies were conducted to investigate the biological response. Recently, a careful analysis of these studies was successfully performed in \cite{R19}, in which the authors explored the existence of THz related effects on gene expression that can be unambiguously distinguished from thermal effects. It was suggested that THz radiation may affect gene expression by perturbing the conformational dynamics of double-stranded DNA (dsDNA) \cite{Alex_PRA,Alex_PRE,Alex_PLoSCB}. These studies were inspired by prior ones \cite{Alexandrov_PLoSCB,Alexandrov_NR}. As THz photons do not carry enough energy to directly alter chemical reactions, nonlinear resonance effects may cause local changes of breathing dynamics in these systems \cite{blank,fisher}.

Motivated by this fact, Alexandrov et al. studied the influence of a THz field into the dynamics of a homogeneous poly(A) DNA molecule with 64 base pairs (bps) \cite{Alex_PRA,Alex_PRE}. To model the interactions of dsDNA with THz field, they made use of the Peyrard-Bishop-Dauxois (PBD) model \cite{PBD}. They regarded periodic driving and frictional terms in the absence of thermal noise. In that study, they found breather modes (localized periodic motions of the double strand) under certain conditions. Hence, they concluded that the main effect of THz radiation is to influence resonantly into the dynamical stability of the dsDNA. Though Swanson later showed that these breather modes can be eliminated by changing the PBD model parameters or by including thermal noise \cite{Swanson}, he agreed that under assumptions concerning drag and drive forcing, breather modes can be generated at certain resonant frequencies.

The PBD model is useful because it allows us to study the equilibrium and dynamic properties of DNA. It has potential ability to describe the melting transition and denaturation bubbles of the dsDNA such as those that occur during the initial stage of the transcription process. Some specific correlation with transcription initiation sites has been claimed \cite{Choi,Kalosakas04,Alexandrov_PLoSCB,Choi1,Alexanrov_pt} and debated \cite{Erp1,reply}.

Tapia et al. \cite{Tapia} set suitable parameter values for PBD model and studied the formation and stabilities of bubbles in the system. They used an enhanced model that includes solvent interactions through the addition of a Gaussian barrier to the Morse potential \cite{Weber}. This barrier modifies the melting transition and the dynamics of the molecule. They also focused on the application of the Principal Component Analysis (PCA) of the trajectories under equilibrium conditions.

Even if both the PBD model and the interaction with external field taken as Alexandrov are quite simple, they could give some insights in the understanding of the interaction of THz field with the DNA molecule. For these reasons our purpose is to study the effect of the THz field on thermal and dynamic properties of DNA: melting transition and bubble formation at physiological temperatures in the framework of the PBD model. We used a modified version of the model with the inclusion of solvation barrier and included thermal noise. Both homogeneous and heterogeneous chain are studied in order to approach more to reality. The heterogeneous chain is the adeno-associated viral P5 promoter (AAV P5 promoter or P5 promoter), which has been widely studied \cite{Kalosakas04,Alexandrov_PLoSCB} and plays an important role in the AAV DNA replication \cite{Virol} and the regulation of the AAV gene expression \cite{Virol1}.

This work is divided as follows: in section II we describe the model; in section III the methods are summarized; in section IV an analysis of the response of the system at different frequencies and field amplitudes is made. After that, we will show the influence of THz field, with specific parameters, into melting transition (section V) and denaturation bubbles (section VI). The last section is devoted to conclusions.

\section{THE MODEL}
The PBD model is a mesoscopic dynamical model of the DNA molecule. It describes the stretching of the bonds between the bps through a single variable, which condenses all the atomic coordinates of a bp. This model ignores the helicoidal structure and uses the Morse potential to model hydrogen bonding between bps. A non-linear inter-pair stacking potential is also considered. We use a modification of the PBD model including a solvation barrier inside of the Morse potential. This barrier prevents the closing of the base once it is opened. The total energy of the system is then approached by:

\begin{equation}
H=\sum_n \left[ \frac{p_n^2}{2m}+V(y_n)+W(y_n,y_{n-1}) \right].
\label{eq:ham}
\end{equation}

In this equation $V(y_n)$ is an on-site potential that describes the interaction between the two bases of a pair. It is represented by the Morse potential and a Gaussian barrier is added:

\begin{equation}
V(y)=D ({\rm e}^{-\alpha y}-1)^2 + G {\rm e}^{-(y-y_0)^2/b}.
\label{eq:Vmod}
\end{equation}

Parameters $D$, $\alpha$, $G$, $y_0$ and $b$ are sequence dependent. Following \cite{Campa},
in our simulations we will use $D_{CG}=1.5D_{AT}$ and $\alpha_{CG}=1.5\alpha_{AT}$. $D$ is the bp dissociation energy and $\alpha$ sets the amplitude of the potential well. The barrier height is controlled by $G$, its position and its width are given by $y_0$ and $b$ respectively. A reasonable selection for such parameters is $G=3D$, $y_0=2/\alpha$ and $b=1/2\alpha^2$ \cite{Tapia}.
The term $W(y_n,y_{n-1})$ accounts for the stacking interactions and is given by

\begin{equation}
W(y_n,y_{n-1})=\frac{1}{2}K(1+\rho{\rm e}^{-\delta(y_n+y_{n-1})})(y_n-y_{n-1})^2.
\end{equation}

The effect of this term, whose intensity is governed by $\rho$, is to change the effective coupling constant from $K(1+\rho)$ to $K$ when one of the bps is displaced away from its equilibrium position. The parameter $\delta$ sets the scale length for this behavior. Alternatively, inhomogeneous stacking energy can be also considered \cite{Weber_stacking}.

We used the same value parameters of the PBD model with solvation barrier that appear in reference \cite{Tapia}: $D_{AT}=0.05185 eV$, $G_{AT}=0.1556 eV$ , $y_{0AT}=0.5 {\AA}$, $b_{AT}=0.03125 {\AA}^2$, $K=0.03 eV{\AA}^2$, $\rho=3$ and $\delta=0.8 {\AA}^{-1}$.

\section{METHODS}
In order to study the behavior of the system we have performed molecular-dynamics numerical simulations of the Langevin equation:

\begin{eqnarray}
m\frac{\partial^2 y_n}{\partial t^2}+m\gamma\frac{\partial y_n}{\partial t} &=&-\frac{\partial\left [W(y_n,y_{n+1}+W(y_{n-1},y_n)\right]}{\partial y_n}  \nonumber \\ & & -\frac{\partial V}{\partial y_n} +\xi_n(t) + A cos(\omega t),
\label{eq:lang_dyn_cont}
\end{eqnarray}

where $m$ is the mass of the bp, $\gamma$ is the effective damping of the system and $\xi(t)$ accounts for thermal noise, $\langle\xi_n(t)\rangle=0$ and $\langle\xi_n(t)\xi_k(t')\rangle=2m\gamma k_BT\delta_{nk}\delta(t-t')$, $T$ is the bath temperature, $A$ and $\omega$ are the field amplitude and frequency respectively. The equations are numerically integrated using a stochastic Runge-Kutta algorithm \cite{sde1,sde2}. We thermalize during $200 ps$  without field and $200 ps$  with the field before any record is made. Simulations of melting transition are performed using periodic boundary conditions while those of thermal bubbles with fixed boundary conditions as in reference \cite{Tapia}. The P5 promoter is given by the 69 bps:
"5- GTGCCCATTTAGGGTATATATGGCCGAGTGAGCGAGCAGGATCTCCATTTTGACCGCAAATTTGAACG - 3". For methodological issues we analyze homogeneous chains as well. In these cases the chains have the same number of bps as the P5 chain.

To show the influence of field parameters in the system response we use the mean displacement
$\langle y\rangle$ defined as

\begin{equation}
\langle y\rangle = \frac{1}{N t_s} \sum_{n,t}^{N,t_s} y_n (t),
\label{eq:ym}
\end{equation}

where $N$ is the number of bps and $t_s$ is the simulation time. For the melting transition we also measure the mean energy $\langle u\rangle$ as a function of the temperature

\begin{equation}
u = \frac{1}{N t_s} \sum_{n,t}^{N,t_s} \left[ W(y_n - y_{n-1}) + V(y_n)
\right].
\label{eq:energy}
\end{equation}

In order to calculate the opening probability and lifetime of a bubble we followed reference \cite{Alexandrov_PLoSCB}. The probability $P_n(l,tr)$ for the existence of a bubble of certain length $l$ of bps, threshold $tr$ and beginning at $n^{th}$  bp is calculated as:

\begin{equation}
P_n(l,tr)=\frac{1}{t_s}\sum_{a=1}^{q_{n}^{max}(l,tr)}\Delta t[q_n(l,tr)],
\label{eq:prob}
\end{equation}

where $q_n(l,tr)$ counts for the bubbles of duration $\Delta t[q_n(l,tr)]$. The average bubble duration $\tau$ is calculated as the average time of a bubble of a given shape over all occurrences of that bubble,

\begin{equation}
\tau_n=\frac{\sum_{a=1}^{q_{n}^{max}(l,tr)}\Delta t[q_n(l,tr)]}{\sum_{a=1}^{q_{n}^{max}(l,tr)}[q_n(l,tr)]}.
\label{eq:t}
\end{equation}

We can extract information from a large set of data in a multidimensional phase space through the PCA. It allows to reduce the dimensionality of the variable to those that include most of the fluctuations of the original system \cite{jolliffe_book}. From an operational point of view, we have to build the $N\times N$ correlation matrix. So,

\begin{equation}
 C(i,j) = \left< y_i y_j \right> - \left< y_i \right> \left< y_j \right>.
\label{eq:correl}
\end{equation}

The diagonalization of this matrix allows us to obtain an ordered set of eigenvalues ($\lambda_1 > \lambda_2 > \lambda_3 ...$) with their corresponding eigenvectors ($v_1, v_2, v_3 ....$). The amount of fluctuations is given by the eigenvalues. The new coordinates are ordered in such a way that most of the system fluctuations are retained by the few first ones.

\section{SYSTEM RESPONSE FOR DIFFERENT FREQUENCIES AND FIELD AMPLITUDES }
Before performing the simulations for the melting transition and bubble formation, some preliminary steps should be done. First, we look for the frequency values at which maximum responses are obtained for each sequence. These values depend on temperature. At low temperatures they should be in the order of the linear resonances of the system. Nonlinear oscillations become important at intermediate temperatures while above the melting temperature, frequency values belong to those of a Gaussian chain.

\begin{figure}
\includegraphics[width=7.5cm]{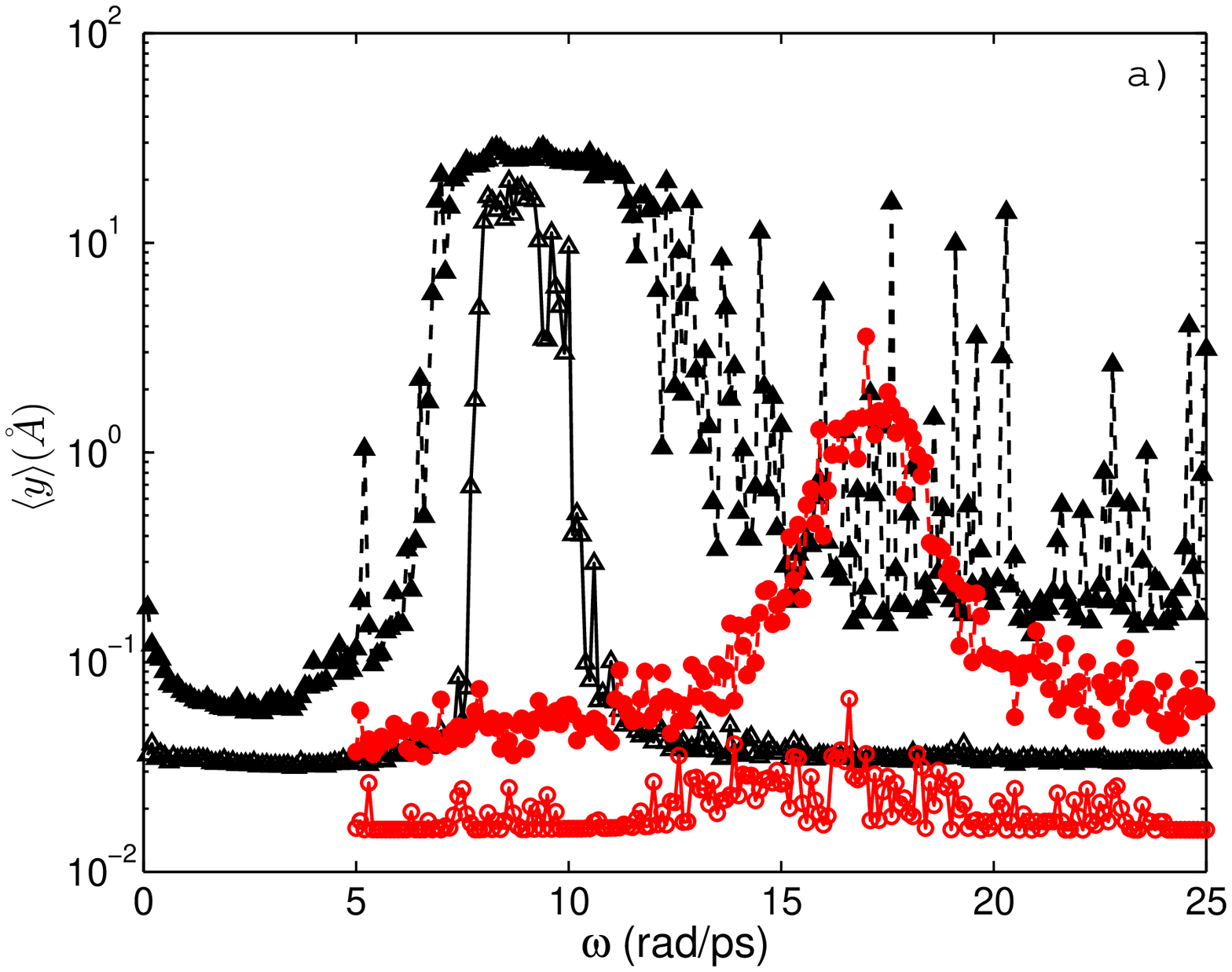}
\includegraphics[width=7.5cm]{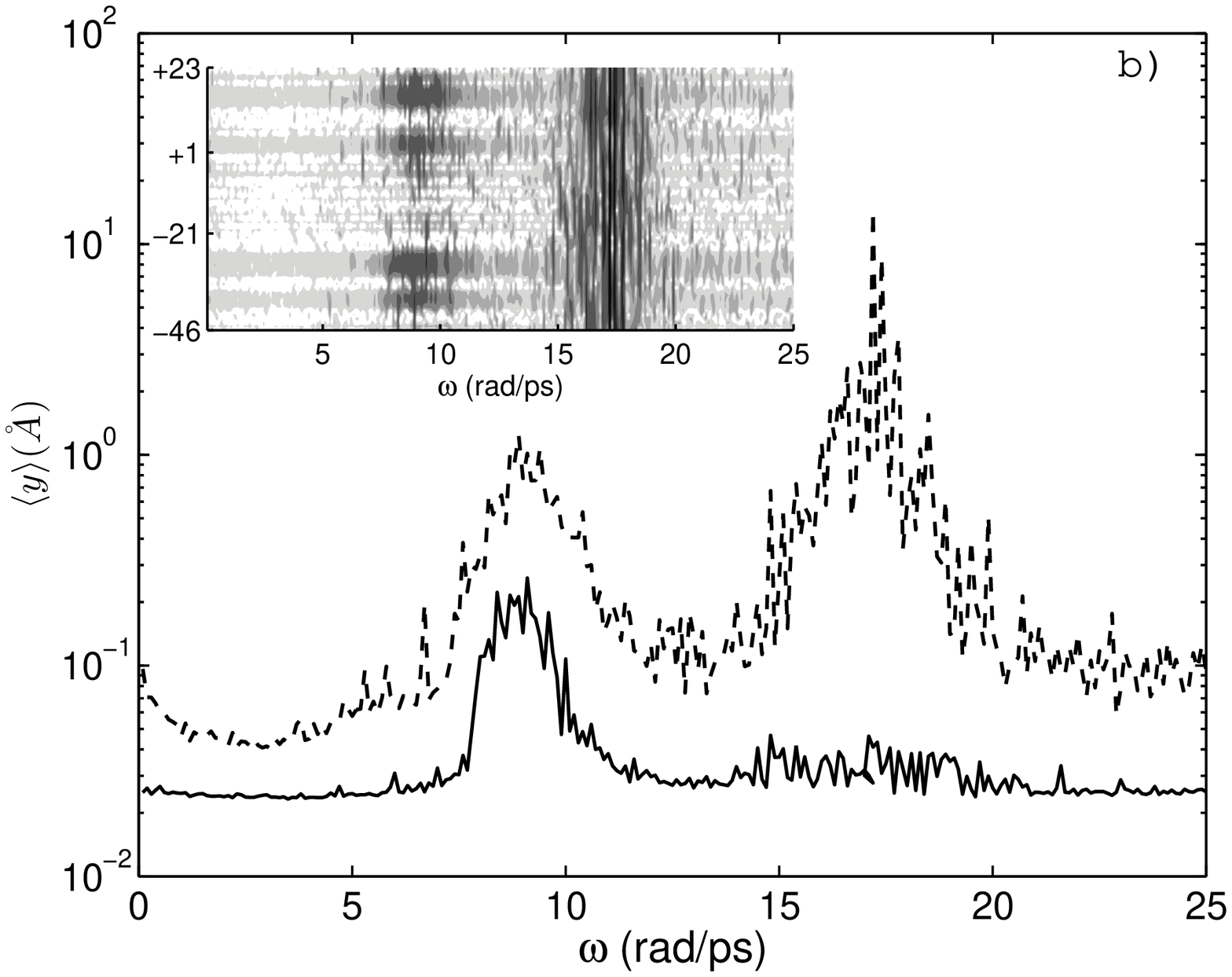}
\caption{(Color online) Frequency dependence of $\langle y\rangle$ with $A=50 pN$ and $\gamma =1 ps^{-1}$.a) AT chain is represented by lines with triangles and CG one by circles. Filled markers are for $T=290 K$ and empty are for $T=210 K$. b) P5 promoter. Solid line is for $T=210 K$ and dashed line for $T=290 K$. Figure inset: Frequency dependence of $\langle y\rangle$ for each base at $T=290 K$.}
\label{fig:ymvsw}
\end{figure}

The system dynamics depends on damping values as well. Thus, we have considered two values for the damping coefficient $\gamma= 1ps^{-1}$ and $\gamma= 9.8 ps^{-1}$.
Figure \ref{fig:ymvsw} shows the behavior of three sequences with 69 bps. The calculation has been performed with a field amplitude $A=50 pN$, damping factor $\gamma =1 ps^{-1}$ and two temperature values: $T=210 K$ and $T=290 K$.

\begin{figure}
\includegraphics[width=7.5cm]{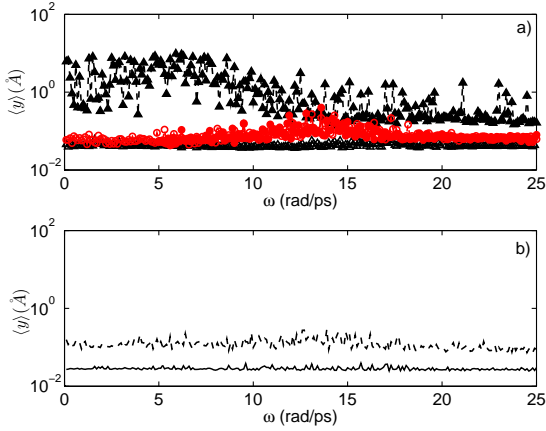}
\caption{(Color online) Frequency dependence of $\langle y\rangle$ with $A=144 pN$ and $\gamma =9.8 ps^{-1}$. a) AT chain is represented by lines with triangles and CG one by circles. Filled markers are for $T=290 K$ and empty are for $T=210 K$. b) P5 promoter. Solid line is for $T=210 K$ and dashed line for $T=290 K$.}
\label{fig:ymvsw1}
\end{figure}

Maximum responses occur at certain frequency values, even for large field amplitudes (not shown). Some modes are activated when temperature increases, (see fig.\ref{fig:ymvsw}). The frequency bandwidth slightly increases when temperature rises. These bands are around $\omega=9 rad/ps$ and $\omega= 17 rad/ps$ for AT and CG chains respectively. The increased of $\langle y\rangle$ for certain frequency values could lead to bubbles formation or the full melting. The resonant frequencies of P5 promoter are the same as those for the AT and CG chains. In other words, P5 promoter behaves as if it would be composed by two homogeneous chains on the frequency space. This behavior may be understood because AT and CG bps number on the chain are approximately equal. However, the $\langle y\rangle$ values are less than the corresponding to the homogenous sequences. At $T=210 K$ the maximum response for P5 promoter occurs around $\omega=9 rad/ps$. Only the AT bases are stimulated and for this reason localized openings are observed around AT richer regions, (see the inset of the fig. \ref{fig:ymvsw} b). At $T=290 K$ the maximum response occurs at both $\omega=9 rad/ps$ and $\omega =17 rad/ps$. The CG bases are also stimulated. In this case, there is enough energy for opening AT bases as well due to stacking interactions and the whole chain opens.
Figure \ref{fig:ymvsw1} illustrates the frequency dependence of $\langle y\rangle$ with $\gamma =9.8 ps^{-1}$. With a larger damping, it is no longer possible to obtain resonant frequencies because the stochastic term becomes dominant.

\begin{figure}
\includegraphics[width=7.5cm]{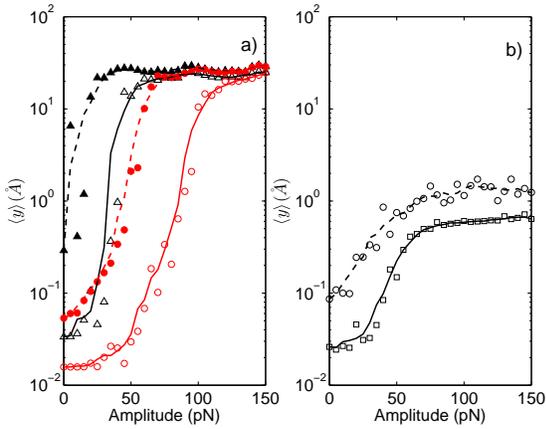}
\caption{(Color online) Amplitude dependence of $\langle y\rangle$ with $\gamma=1 ps^{-1}$. a) AT chain is represented with triangles and CG one with circles. Filled markers are for $T=290 K$ and empty are for $T=210 K$. b) P5 promoter. Squares are for $T=210 K$ and circles for $T=290 K$. Lines are guide to eyes.}
\label{fig:ymvsA}
\end{figure}

\begin{figure}
\includegraphics[width=7.5cm]{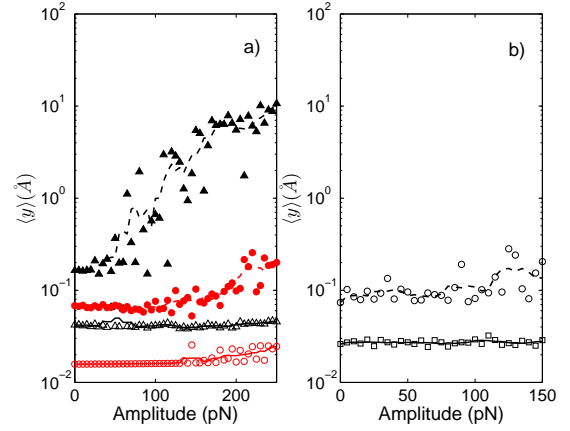}
\caption{(Color online) Amplitude dependence of $\langle y\rangle$ with $\gamma=9.8 ps^{-1}$. a) AT chain is represented with triangles and CG one with circles. Filled markers are for $T=290 K$ and empty are for $T=210 K$. b) P5 promoter. Squares are for $T=210 K$ and circles for $T=290 K$. Lines are guide to eyes.}
\label{fig:ymvsA1}
\end{figure}

We have also calculated the amplitude dependence of $\langle y\rangle$ to determine $A$ values for the next simulations. The frequency values of maximum responses are used: $\omega=9 rad/ps$ for AT and P5 chains and $\omega=17 rad/ps$ for CG chain. Results for $\omega=17 rad/ps$ for the P5 chain are similar to those of $\omega=9 rad/ps$. Figures \ref{fig:ymvsA} and \ref{fig:ymvsA1} show the results for $\gamma= 1 ps^{-1}$ and $\gamma= 9.8 ps^{-1}$ respectively.

According to fig. \ref{fig:ymvsA} for $\gamma= 1ps^{-1}$, we can use $A=10, 25$ and $50 pN$ for the three chains. In the case of $\gamma= 9.8ps^{-1}$ the values are $A=50, 144$ and $200 pN$.

\section{Melting Transition}
\begin{figure}
\includegraphics[width=7.5cm]{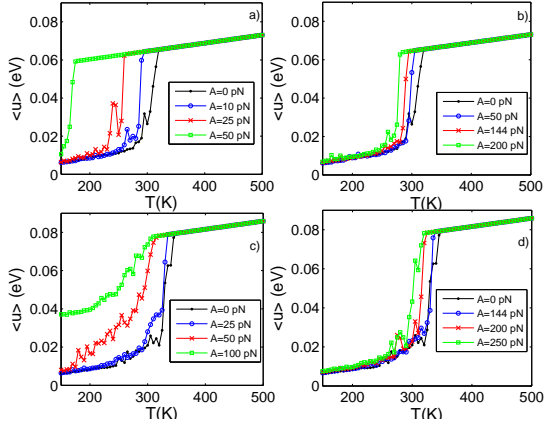}
\caption{(Color online) Melting transition at the frequency $\omega =9 rad/ps$. a) AT chain with damping factor $\gamma=1 ps^{-1}$, b) AT chain with damping factor $\gamma=9.8 ps^{-1}$, c) P5 chain with damping factor $\gamma=1 ps^{-1}$ and d) P5 chain with damping factor $\gamma=9.8 ps^{-1}$.}
\label{fig:transition}
\end{figure}

This section focuses on the study of the melting transition of the homogeneous chains and P5 promoter. Our goal is to analyze how a THz field modifies melting temperature $T_m$ and the transition width $\Delta T$. These parameters were determined in reference \cite{Tapia} for the uniform chain of AT bps without external field. Following the same criteria we can determine them when the external field is applied. We calculate the mean potential energy and mean displacement as a function of the temperature. Let us determinate two temperatures as follow:  the temperature $T_1$ estimates the beginning of the transition where $\langle y(T)\rangle$ crosses 0.5 ${\AA}$. The larger one, $T_2$, provides the onset of the linear behavior in $\langle y(T)\rangle$. Both quantities are defined in the following form: $T_m=(T_1+T_2)/2$ and $\Delta T=(T_2-T_1)/2$.
Figure \ref{fig:transition} displays melting transition curves for the AT chain and P5 promoter. Melting transitions curves for CG and P5 chains at $\omega= 17 rad/ps$ have similar behaviors (not shown).

\begin{figure}
\includegraphics[width=7.5cm]{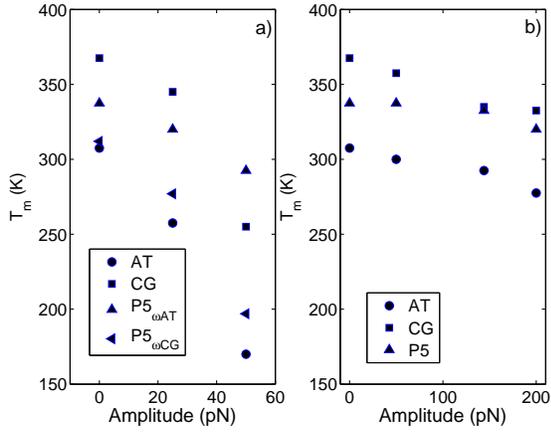}
\caption{Melting transition temperature. a) damping factor $\gamma=1 ps^{-1}$. frequency $\omega =9 rad/ps$ for AT and P5 chains $(P5_{\omega AT})$ and $\omega =17 rad/ps$ for CG and P5 chains $(P5_{\omega CG})$., b) damping factor $\gamma=9.8 ps^{-1}$. frequency $\omega =9 rad/ps$ for AT and P5 chains and $\omega =17 rad/ps$ for CG chain.}
\label{fig:valores}
\end{figure}

While the mean potential  energy curves coincide above $T_2$ for different field amplitude and damping values, the mean displacement curves differ. This can be explained because the action of the field is such that the difference $(y_n-y_{n-1})$ between two successive bps at each time step remains constant. In this region the potential energy depends only on the difference $(y_n-y_{n-1})$. Figure \ref{fig:valores} shows the behavior of the melting transition temperature versus field amplitude for the three chains.

The differences between the melting transition with and without applied field are remarkable (see figs. \ref{fig:transition} and \ref{fig:valores}). Due to the applied external field, the chains melt at lower temperatures. This fact has been already noted by Swanson \cite{Swanson} for a homogeneous chain. Differently from this work, the values of $T_m$ we obtained here are larger than the one there reported. Behavior of the transition width is more complex because it depends on the number of bps that has been opened at certain temperature. The field allows both opening and closing of bps. For $\gamma=9.8 ps^{-1}$ the effect of the field decreases and larger values of field amplitude are needed to lower the transition temperature.

\section{Bubbles Formation}
\begin{figure}
\includegraphics[width=7.5cm]{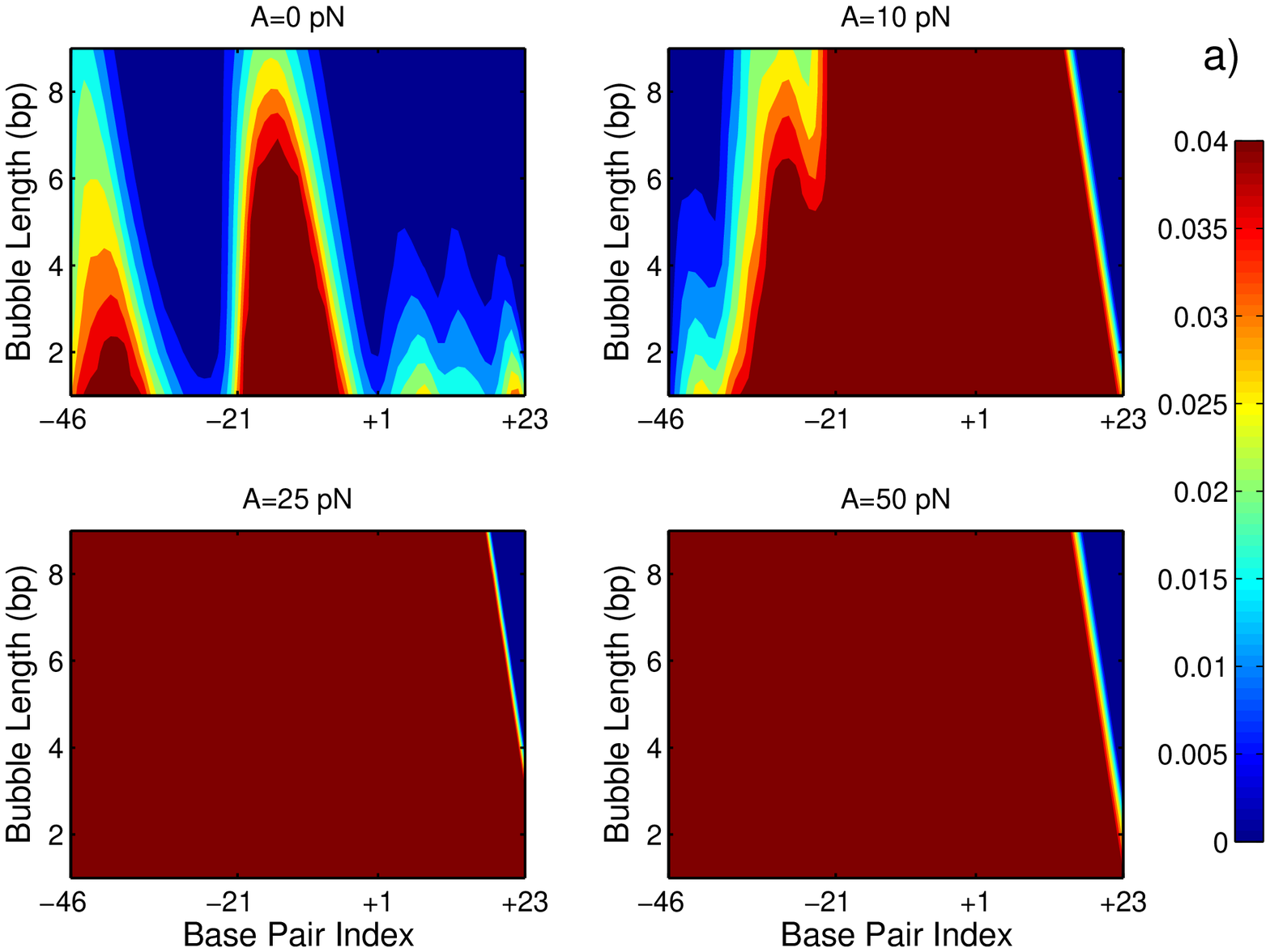}
\includegraphics[width=7.5cm]{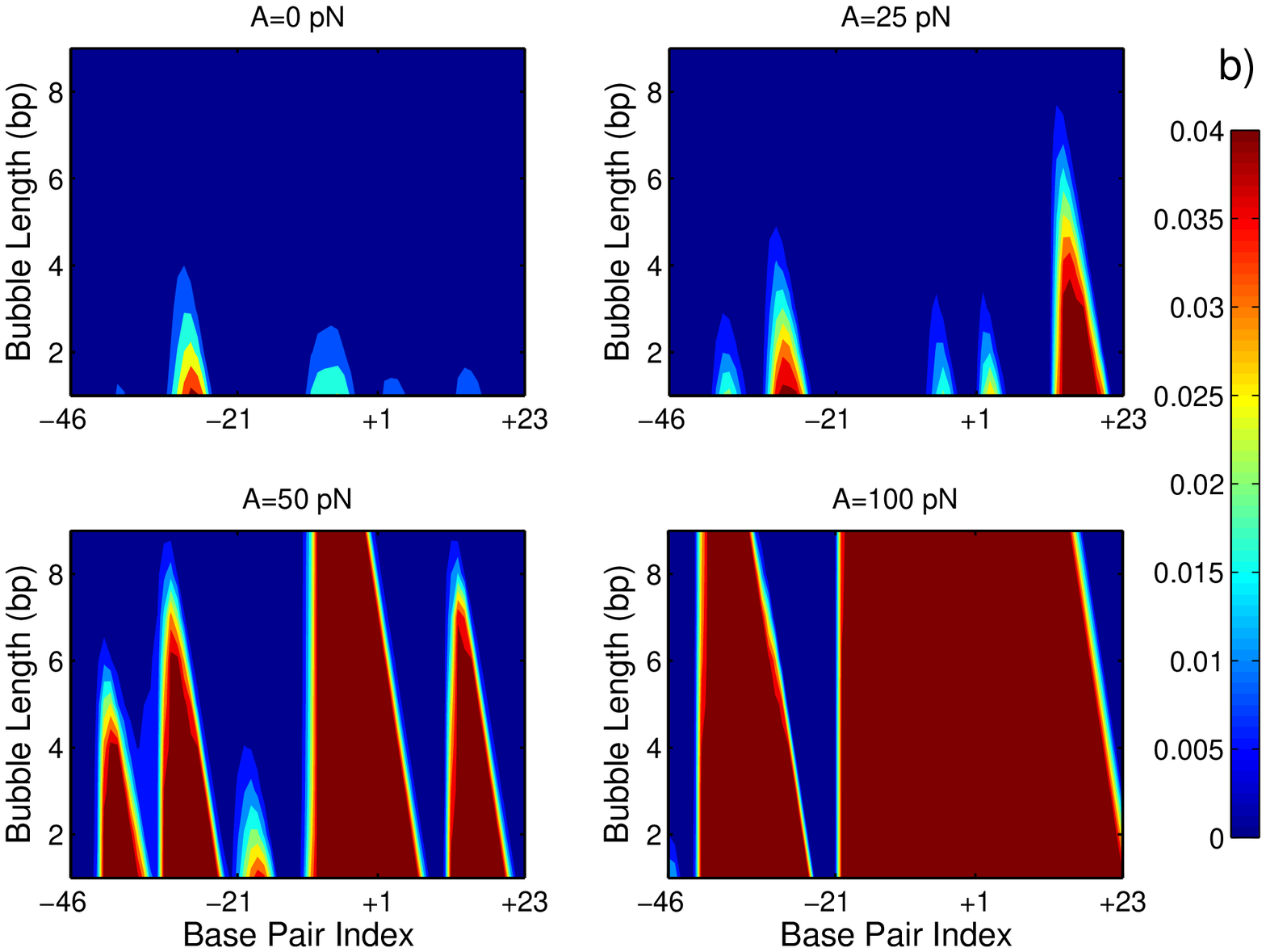}
\caption{(Color online) Probability opening distribution $P_n$ with frequency $\omega=9 rad/ps$, damping factor $\gamma= 1ps^{-1}$ and $T=290 K$. a) AT chain b) P5 promoter.}
\label{fig:Pn1}
\end{figure}

\begin{figure}
\includegraphics[width=7.5cm]{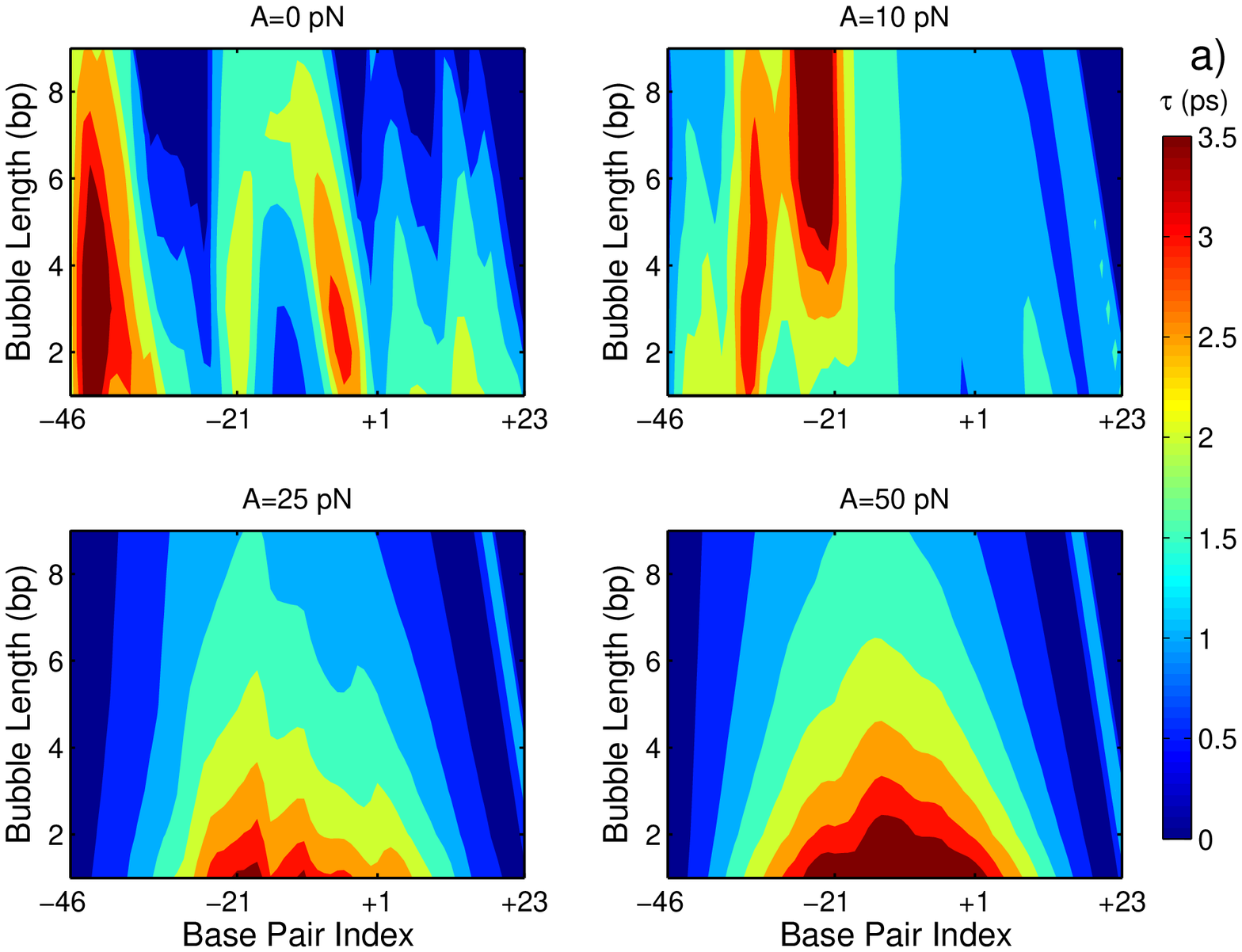}
\includegraphics[width=7.5cm]{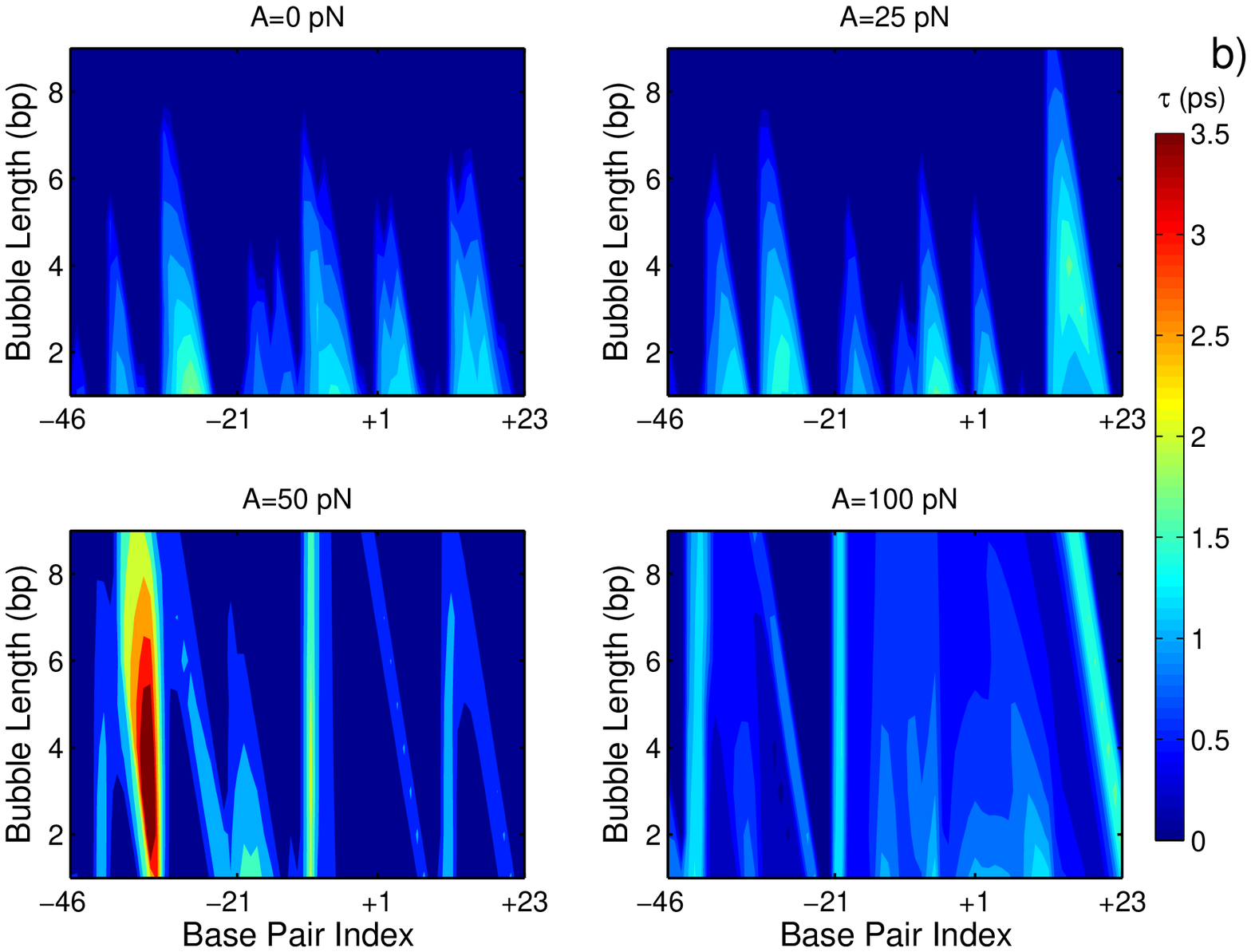}
\caption{(Color online) Average lifetime distribution $\tau_n$ with frequency $\omega=9 rad/ps$, damping factor $\gamma= 1ps^{-1}$ and $T=290 K$. a) AT chain b) P5 promoter.}
\label{fig:tn1}
\end{figure}

Next, we study the THz field influence in the bubbles formation  at $T=290 K$ with the parameters chosen in previous section. This temperature belongs to the premelting range, in which it has been shown that the highest opening probability coincides with important biological sites like the start transition site (TSS) and the TATA box \cite{Choi,Kalosakas04,Choi1}. In order to avoid unphysical denaturation process due to finite-size effects, we add 10 CG bps sequence to the ends of the P5 promoter to create hard boundaries. The extremes are set to zero , avoiding the complete opening of the chain \cite{Tapia}.
Figures \ref{fig:Pn1} and \ref{fig:tn1} show the bubble size and lifetime distribution respectively for the AT chain and P5 promoter with $\omega=9 rad/ps$ and $\gamma= 1ps^{-1}$. $P_n$ and $\tau_n$, given by equations \ref{eq:prob} and \ref{eq:t}, are defined with a threshold value of $tr=1.5 {\AA}$. These magnitudes are represented as a function of the bubble length and index site. The opening probability and bubbles lifetime are given in color scale. In these figures the 10 bps at the beginning and the end of the sequence are not included. The +1 in Base Pair Index refers to TSS position in P5 promoter. In the homogeneous chain there is no TSS but we keep the same notation for convenience. 

\begin{figure}
\includegraphics[width=7.5cm]{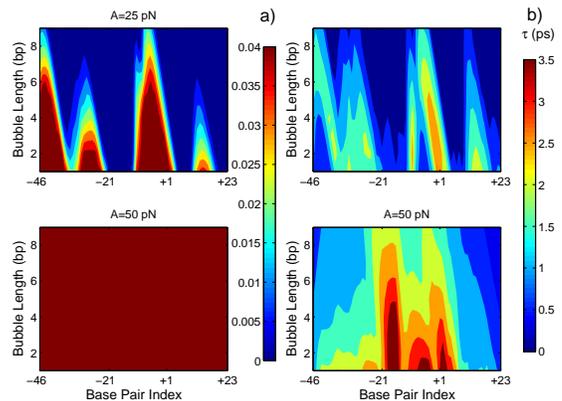}
\caption{(Color online) Probability opening distribution $P_n$ and average lifetime distribution $\tau_n$ for P5 promoter with frequency $\omega=17 rad/ps$, damping factor $\gamma= 1ps^{-1}$ and $T=290 K$. Top figure: $A= 25 pN$. Bottom figure: $A= 50 pN$.}
\label{fig:w17}
\end{figure}

Figure \ref{fig:Pn1} shows how the external field enhances the opening probability at the frequency value $\omega=9 rad/ps$. Results agree with those of the melting transition. Without field, the larger probabilities in P5 promoter occur in two sites of biological interest as previously reported: the TSS (represented by +1) and the TATA box (between -30 and -40). Increasing the field amplitude leads to the increase of opening probability at these sites and helps the opening of others. Without the THz field, the opening probability of P5 promoter at the TATA box is higher than at the TSS \cite{Alexanrov_pt,ERp_2}.
The most persistent bubbles are found at the sites that have been pointed out before, see Fig. \ref{fig:tn1}. In our simulations, the lifetime values depend on the selection of the parameters of the modified PBD model. Thus, the results in some cases can be different respect to those reported in literature when the THz field is not applied. For the AT chain, the field makes bubbles more stable but for the P5 promoter it does not. In this case, the heterogeneity of the chain plays a crucial role, because we have used $\omega=9 rad/ps$. The decrease of bubbles lifetime may be explained because the energy of the field favors both opening and closing events of the bps. The opening and closing kinetics is governed by the solvation barrier and the applied field. With barrier and without applied field, the kinetics is controlled by the presence of two equilibrium states separated by the solvation barrier. Closing events are more difficult and bubbles live longer than those in which the barrier is not considered \cite{Tapia}. In the other hand, at this frequency CG bps are not stimulated. If we use $\omega=17 rad/ps$, CG bps are stimulated and the whole chain opens, as explained previously, Fig. \ref{fig:w17}. The same calculations were performed with $\gamma=9.8 ps^{-1}$ (not shown). As in previous results, damping factor modifies the chain dynamics and high field amplitude values are needed to open the chain.

\begin{figure}
\includegraphics[width=7.5cm]{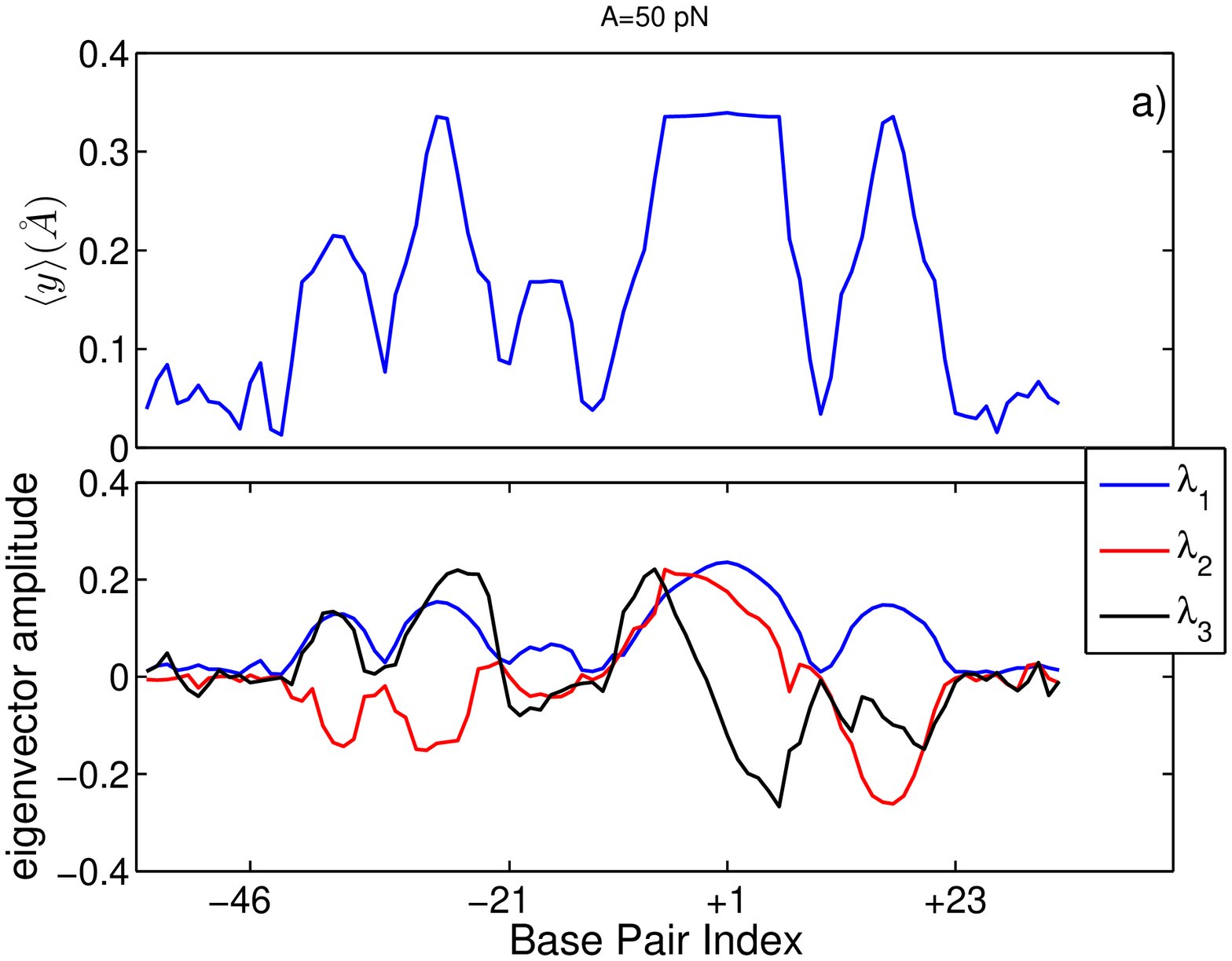}
\includegraphics[width=7.5cm]{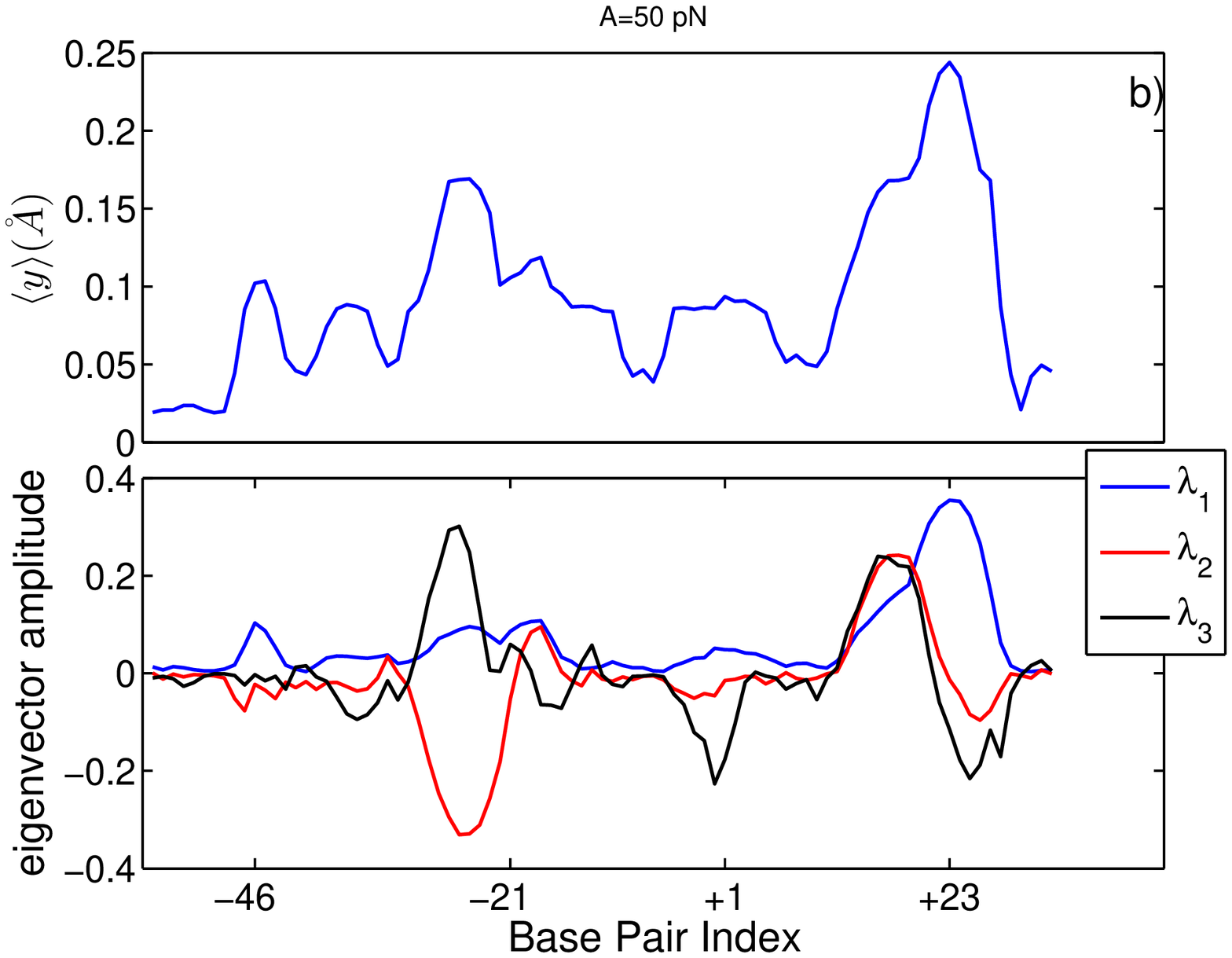}
\caption{(Color online) Top figure: probability of opening for P5 promoter. Bottom figure: the three
first PC eigenvectors corresponding to eigenvalues $\lambda_1$, $\lambda_2$ and $\lambda_3$ at $T=290 K$. a) $\gamma= 1 ps^{-1}$. b) $\gamma= 9.8 ps^{-1}$.}
\label{fig:PCA}
\end{figure}

Finally, we compare the results obtained from the PCA with the average displacement values of each base (see figure \ref{fig:PCA}). The 10 CG bps at the ends of the sequence are included in the figures. We only show the results for P5 promoter with a field amplitude of $A=50 pN$.

A good correlation between top figures and bottom ones is verified. Localized eigenvectors span over regions of nine bps, which fairly correspond to the width of the bubbles, as in reference \cite{Tapia}. For other field amplitudes and the homogeneous chains there is a good correlation too.

Our results suggest that the THz field influence can be viewed in two main directions. First, the applied field tends to facilitate melting and bubble formation which could in principle affects processes like transcription or replication. The driving forces needed for observing such effects are large compared with those found in physical reliable conditions for in vivo exposure \cite{Swanson}. We do not disagree with this conclusion. We have just used a simple model for such a description and the magnitudes values may no match the real ones. On the other hand, the external field could be used to detect biologically significant sites by increasing the opening probability at these sites without the melting of the chain. In this scenario the effective drive could be larger.

\section{CONCLUDING REMARKS}

We have studied in the framework of PBD model the influence of THz field on homogeneous chains and the heterogeneous P5 promoter. Thermal properties of these sequences have been studied by including thermal noise and a solvation barrier in the model. 
The influence of THz field depends strongly on field parameters (frequency, field amplitude) and system parameters (potential parameters, damping). In spite of previous results, we do not obtain breathers modes (oscillatory solutions) but rather we find that ac-fields favors the formation of bubbles. 

We have also identified the frequency resonant bands that mostly increase the opening of the chain. The position of the  bands are sequence dependent and distinguish the AT-rich from CG-rich regions. This could increase experimental resolution in order to detect sites like TSS if small field amplitudes are used. More study in this direction needs to be done, for instance considering more complex interaction between bases and the external field.

Finally, we have numerically obtained that the PCA can be also used to get information for out equilibrium systems.

\begin{acknowledgments}
We thank R. Tapia-Rojo for assistance with model parameters and a critical reading of the manuscript. The work has been supported by the 
Spanish Project No. FIS2011-25167 co-financed by Fondo Europeo de Desarrollo Regional (FEDER) funds.

\end{acknowledgments}


\begin{thebibliography}{99}

\bibitem{R2} R. Appleby and H. B. Wallace, IEEE Trans. Antenn. Propag {\bf 55}, 2944 (2007).

\bibitem{R3} J. F. Federici, B. Schulkin, F. Huang, D. Gary, R. Barat, F. Oliveira and D. Zimdars, Semicond. Sci. Technol. {\bf 20}, 266 (2005).

\bibitem{R4} D. Zimdars and J. S. White, Proc. SPIE {\bf 5411}, 78-83 (2004).

\bibitem{R5} R. Bogue, Sensor Review {\bf 29}, 6 (2009).

\bibitem{R6} A. Dobroiu, C. Otani and K. Kawase, Measurement Science and Technology {\bf 17}, 161 (2006).

\bibitem{R7} P. C. Ashworth, E. Pickwell-MacPherson, E. Provenzano, S. E. Pinder, A. D. Purushotham, M. Pepper, and V. P. Wallace, Opt Express {\bf 17}, 12444 (2009).

\bibitem{R8} M. A. Brun, F. Formanek, A. Yasuda, M. Sekine, N. Ando, and Y. Eishii, Phys. Med. Biol. {\bf 55}, 4615 (2010).

\bibitem{R9} S. J. Oh, J. Kang, I. Maeng, J. S. Suh, Y. M. Huh, S. Haam, and J. H. Son, Opt. Express {\bf 17}, 3469 (2009).

\bibitem{R10} V. P. Wallace, P. F. Taday, A. J. Fitzgerald, R. M. Woodward, J. Cluff, R. J. Pye, and D. D. Arnone, Faraday Discuss {\bf 126}, 255 (2004).

\bibitem{R11} Z. D. Taylor, R. S. Singh, M. O. Culjat, J. Y. Suen, W. S. Grundfest, H. Lee, and E. R. Brown, Opt. Lett. {\bf 33}, 1258 (2008).

\bibitem{R12} R. M. Woodward, B. E. Cole, V. P. Wallace, R. J. Pye, D. D. Arnone, E. H. Linfield, and M. Pepper, Phys. Med. Biol. {\bf 47}, 3853 (2002).

\bibitem{R13} P. U. Jepsen, U. Moller, and H. Merbold, Opt. Express {\bf 15}, 14717 (2007).

\bibitem{R14} L. Thrane, R. H. Jacobsen, P. U. Jepsen, and S. R. Keiding, Chem. Phys. Lett. {\bf 240}, 330(1995).

\bibitem{R15} G. J. Wilmink and J. E. Grundt, J. Infrared Milli Terahz Waves {\bf 32}, 1074 (2011).

\bibitem{R16} P. Weightman 2007 Proc. 32nd Int. Conf. on Infrared and Millimeter Waves pp 1-3 IEE number 07EX1863C

\bibitem{R17} P. Weightman, Phys. Biol. {\bf 9}, 053001 (2012).

\bibitem{R18} THz-BRIDGE Project web-page (2004), http://www.frascati.enea.it/THz-BRIDGE/.

\bibitem{R19} B. S. Alexandrov, K. O. Rasmussen, A. R. Bishop, A. Usheva, L. B. Alexandrov, S. Chong, Y. Dagon, L. G. Booshehri, Ch. H. Mielke, M. L. Phipps, J. S. Martinez, H. T. Chen, and G. Rodriguez, Biomedical Optics Express {\bf 9}, 2679 (2011).

\bibitem{Alex_PRA} B. S. Alexandrov, V. Gelev, A. R. Bishop, A. Usheva, and K. O. Rasmussen, Phys. Lett. A {\bf 374}, 1214 (2010).

\bibitem{Alex_PRE} P. Maniadis, B. S. Alexandrov, A. R. Bishop, and K. O. Rasmussen, Phys. Rev. E {\bf 83}, 011904 (2011).

\bibitem{Alex_PLoSCB} J. Bock, Y. Fukuyo, S. Kang, M. L. Phipps, L. B. Alexandrov, K. O. Rasmussen, A. R. Bishop, E. D. Rosen, J. S. Martinez, H. T. Chen, G. Rodriguez, B. S. Alexandrov, and A. Usheva, PLoS ONE  {\bf 5}, 15806 (2010).

\bibitem{Alexandrov_PLoSCB}
B. S. Alexandrov, V. Gelev, S. W. Yoo, A. R. Bishop, K. O. Rasmussen, and A. Usheva, PLoS Computational Biology  {\bf 5}, 1000313 (2009)


\bibitem{Alexandrov_NR}
B. S. Alexandrov, V. Gelev, S. W. Yoo, L. B. Alexandrov, Y. Fukuyo, A. R. Bishop, K. O. Rasmussen, and A. Usheva, Nucleic Acids Res.  {\bf 38},  1790 (2010).

\bibitem{blank}
M. Blank and R. Goodman, Bioelectromagnetics  {\bf 18},  111 (1997).

\bibitem{fisher}
B. M. Fischer, M. Walther, and P. Uhd Jepsen, Phys. Med. Biol.  {\bf 47},  3807 (2002).

\bibitem{PBD} T. Dauxois, M. Peyrard, and A. R. Bishop, Phys. Rev. E {\bf 47}, 684 (1993).

\bibitem{Swanson} E. S. Swanson, Phys. Rev. E {\bf 83}, 040901 (2011).

\bibitem{Choi}
C. H. Choi, G. Kalosakas, K. O. Rasmussen, M. Hiromura, A. R. Bishop, A. Usheva, Nucleic Acid Research {\bf 32}, 1584 (2004).

\bibitem{Kalosakas04}
G. Kalosakas, K. O. Rasmussen, A. R. Bishop, C. H. Choi, A. Usheva,
Eur. Phys. Lett. {\bf 68}, 127 (2004)

\bibitem{Choi1}
C. H. Choi, Z. Rapti, V. Gelev, M. R. Hacker, B. S. Alexandrov, E. J. Park, J. S. Park, N. Horikoshi, A. Smerzi, K. O. Rasmussen, A. R. Bishop, and A. Usheva, Biophys J. {\bf 95}, 597 (2008).

\bibitem{Alexanrov_pt}
B. S. Alexandrov, L. T. Wille, K. O. Rasmussen, A. R. Bishop, K. B. Blagoev, Phys. Rev. E {\bf 74}, 050901 (2006).

\bibitem{reply}
C. H. Choi, A. Usheva, G. Kalosakas, K. O. Rasmussen and A.R. Bishop,
Phys. Rev. Lett.,{\bf 96}, 239801 (2006); Reply by T. S. van Erp, S. Cuesta-Lopez, J-G. Hagmann and M. Peyrard, Phys. Rev. Lett. {\bf 96}, 239802 (2006).

\bibitem{Erp1} T. S. van Erp, S. Cuesta-Lopez, and M. Peyrard,
Eur. Phys. J. E {\bf 20}, 421 (2006).

\bibitem{Tapia} R. Tapia-Rojo, J. J. Mazo, and F. Falo, Phys. Rev. E {\bf 82}, 031916 (2010).

\bibitem{Weber} G. Weber, Phys. EPL {\bf 73}, 806 (2006).

\bibitem{Virol}
J. D. Tratschin, I. L. Miller, and B. J. Carter, J. Virol.  {\bf 51},  611 (1984).

\bibitem{Virol1}
M. A. Labow, P. L. Hermonat, and K. I. Berns, J. Virol.  {\bf 60},  251 (1984).

\bibitem{Campa} A. Campa and A. Giansanti, Phys. Rev. E {\bf 58}, 3585 (1998).

\bibitem{Weber_stacking} G. Weber, J. W. Essex, and C. Neylon, Nature Physics {\bf 5}, 769 (2009).

\bibitem{jolliffe_book} I. T. Jolliffe, {\it Principal Components Analysis 2ed}
(Springer-Verlag New York 2002).

\bibitem{sde2} H. S. Greenside and E. Helfand, The Bell System Technical Journal {\bf 60}, 1927 (1981).

\bibitem{sde1} E. Helfand,
Bell System Technical Journal {\bf 58}, 2289 (1979).

\bibitem{ERp_2}
T. S. van Erp, S. Cuesta-Lopez, J-G. Hagmann and M. Peyrard,
Phys. Rev. Lett. {\bf 95}, 218104 (2005).

\end{thebibliography}
\end{document}